\documentclass{iau}
\usepackage{refjournaux}
\usepackage{graphicx}

\title[ ]
{The dichotomy between strong and ultra-weak magnetic fields among intermediate-mass stars} 

\author[F. Ligni\`eres et al.]   
{Fran\c{c}ois Ligni\`eres$^{1,2}$,
Pascal Petit$^{1,2}$,
Michel Auri\`ere$^{1,2}$,
Gregg A. Wade$^{3}$,
\and Torsten B\"ohm$^{1,2}$}	

\affiliation{$^1$CNRS, Institut de Recherche en Astrophysique et Plan\'etologie \\
14 avenue Edouard Belin, 31400 Toulouse, France
\\ email: {\tt francois.lignieres@irap.omp.eu} \\[\affilskip]
$^2$Universit\'e de Toulouse, UPS-OMP, IRAP \\
31400 Toulouse, France\\
$^3$ Department of Physics, Royal Military College of Canada\\
PO
Box 17000, Station Forces, Kingston, Ontario K7K 7B4, Canada}

\pubyear{2008}
\volume{xxx}  
\pagerange{119--126}
\jname{Magnetic fields throughout stellar evolution}
\editors{A.C. Editor, B.D. Editor \& C.E. Editor, eds.}
\begin{document}

\maketitle

\begin{abstract}
Until recently, the detection of magnetic fields
at the surface of intermediate-mass main-sequence stars has been limited to Ap/Bp stars, a class of    
chemically peculiar stars. This class represents no more than 5-10\% of the stars in this mass range.
This small fraction is not explained by the fossil field paradigm that describes the Ap/Bp type magnetism as a remnant of an early
phase of the star-life. Also, the limitation of the field measurements to a small and special group of stars
is obviously a problem to study the effect of the magnetic fields on
the stellar evolution of a typical intermediate-mass star.
 

Thanks to the improved sensitivity of a new generation of spectropolarimeters, a lower bound
to the magnetic fields of Ap/Bp stars, a two orders of magnitude
desert in the longitudinal magnetic field and a new type of sub-gauss magnetism first discovered on Vega have been identified.
These advances provide new clues to
understand 
the origin of intermediate-mass magnetism as well as its influence on stellar evolution.
In particular, a scenario has been proposed whereby the magnetic dichotomy between Ap/Bp and Vega-like
magnetism 
originate from the bifurcation between stable
and unstable large scale magnetic configurations in differentially rotating stars.
In this paper, we review these recent observational findings
and discuss this scenario.

\end{abstract}

\firstsection 
\section{Introduction}

The origin of the stellar magnetic fields and their effects on the star structure and evolution are the two basics
questions of stellar magnetism.
In all stars that possess a convective envelope, a magnetic field is believed to be generated by dynamo mechanism.
It is not always possible to detect it, but its presence at the surface of these solar-type stars makes little doubt.
For these stars, we also know that a magnetized wind exerts a breaking torque that strongly affects their angular momentum evolution.
This illustrates that although limited our level of understanding of the magnetic fields of solar-type stars
enables to answer at least some simple questions
about their origin and their impact.
This is not the case for hotter stars with radiative envelope where the origin of the observed fields as well as their impact on the star evolution remain largely mysterious.
The properties of the observed fields
are
nevertheless well established. They are large scale mostly dipolar fields with dipole strength ranging from 300 G to 30 kG and are 
remarkably stable over time (Donati \& Landstreet, 2009).
These fields are usually called fossil fields as a reference to the hypothesis that
describes them as remnant of an early phase of the star-life, either from
the collapse of a magnetized cloud or during the convective protostellar phase.
But 
they only 
are detected in a small 5-10\% fraction of the intermediate-mass and massive stars
and this fact did not receive a convincing explanation yet.
From the point of view of stellar evolution, such a lack of understanding of the field generation processes combined with the absence of direct measurements
for the vast majority (90-95 \%) of stars are a major obstacle to model 
the effect of a magnetic field on the structure and
evolution of typical intermediate-mass and massive stars.


In this paper, we review recent observational results that shed a new light on the 5-10\% problem 
and suggested new scenarios for the origin of hot star magnetism.
These results
have been obtained thanks
to the high sensitivity 
of a new generation of
high-resolution \'echelle spectropolarimeters (MuSiCoS, Narval, ESPaDOnS, HARPSpol). In particular, 
their large wavelength coverage 
allows to sum up the magnetically polarized signal 
of many spectral lines and thus to detect weak fields.
The number of lines available being an important factor, it is comparatively easier to detect weak fields in cooler stars.
For example, \cite{AK09} showed that a sub-gauss longitudinal field can be detected on the bright late-type star Pollux 
averaging only a few high S/N Stokes V of Narval or ESPaDOnS.
For the same reason, weak fields are easier to detect in intermediate-mass stars than in more massive stars.

Spectropolarimetric surveys of Ap stars as well as of A-type non-Ap stars have first ruled out the possibility that the 5-10\%
fraction simply reflects the detection limit of magnetic measurements. Indeed, the magnetic fields of Ap/Bp stars were found to 
be higher than a lower limit of the order of 100 G for the longitudinal field while no field were detected in A-type non-Ap stars. 
Then two detections of sub-gauss fields on the bright A stars Vega and Sirius have been obtained. 
Thus, instead of a single class of magnetic stars, intermediate-mass star magnetism is now characterized by
two type of magnetisms, Ap/Bp and Vega-like, separated by a two orders of magnitude magnetic desert between 1G - 100 G.
In section 2, we present the evidences for the lower bound to the Ap/Bp magnetic fields
and the magnetic desert, then the discovery of a new type of sub-gauss magnetism
is described in section 3. The implications for the origin of upper-main-sequence magnetism 
are discussed in section 4.

\section{The lower bound of Ap/Bp magnetic fields and the magnetic desert}



From the first Zeeman effect measurement (Babcock 1947) to the early 90s, all magnetic field detections in the upper-main-sequence have been achieved among
the Ap/Bp stars, a group of late B, A and early F stars showing
strong chemical peculiarities. 
The vector modulus measured from the line broadening in the intensity spectrum ranged from  2 and 30 kG
while measurements of the circular
polarization of spectral lines allowed to find weaker fields, down to $\sim 300$ G dipolar fields.


At that time, the detection limit was too high to study Ap/Bp stars with weak magnetic fields
or to put strong constraints on the upper bound of the magnetic fields in A-type non Ap/Bp stars (Landstreet 1982).
Thus, although this was suspected, it was not possible to confirm that all Ap/Bp stars were magnetic.
Moreover a low-field continuation of the Ap/Bp magnetism among non-Ap/Bp stars could not be ruled out.

This last point is neither excluded from our understanding of the link between
the chemical anomalies and magnetic fields in Ap/Bp stars.
Michaud et al. (1970, 1976) indeed proposed that the magnetic fields of Ap/Bp stars are strong enough to avoid
 macroscopic mixing in the envelope either through 
differential rotation, thermal convection or stellar winds. He
showed
that in the quiescent outer layers microscopic diffusion processes
produce strong anomalies in the chemical abundances
compatible with those observed at the surfaces of Ap/Bp stars.
This model also explains why chemical anomalies are either absent (in normal stars) or reduced (in the slowly rotating Am stars)
when no fields are detected.
Following this picture, we can conclude that all Ap/Bp stars should possess a magnetic field higher than the minimum field strength 
required to suppress macroscopic mixing and that, 
if a magnetic field is present at the surface of A-type non-Ap/Bp stars, its strength should be smaller than this minimum field.


From the mid-90s, deep surveys have been conducted with the spectropolarimeters MuSiCoS and Narval at Telescope Bernard Lyot
to explore the weakly magnetic Ap/Bp stars and non Ap/Bp stars.
\cite{AG07} selected a sample of 28 Ap/Bp stars for which previous attempts had led to no detection or to unreliable ones.
Thanks to the improved sensitivity of these instruments, a magnetic field was found in all the stars of the sample confirming that all Ap/Bp stars are magnetic.
Moreover, this survey proved the existence of a lower bound to the magnetic field of Ap/Bp stars. The maximum absolute value reached by
the longitudinal field as the star rotates $B_L^{\mbox{max}}$ is higher than $\sim 100$ G, while the dipolar fields obtained by fitting an oblique rotator model - 
a magnetic dipole
inclined with respect to the rotation axis - is higher than $B_{\rm min} = 300$ G in all but two stars for which the best model is slightly lower but still
compatible with the 300 G value within the error bars. 


This result is fully compatible with the idea that a minimum field strength is needed
to produce the Ap/Bp chemical anomalies.
But one must keep in mind that the observed lower bound of Ap/Bp magnetic fields could be either higher or equal to
this minimum field.
If it is equal, then one would expect to find a low field extension of the Ap/Bp magnetism among the other intermediate-mass stars.
Indeed, it is  difficult to imagine that a mechanism only generates magnetic fields in a sub-group of stars and that, in addition,
the smallest field generated this way is 
exactly the minimum field required for Ap/Bp chemical anomalies.

The simplest extension one could think of is a population of dipolar-like fields that continuously extends 
the distribution of Ap/Bp dipolar fields below 300 G.
However, surveys among A and late-B stars conducted with MuSiCoS, Narval and HARPSpol (Shorlin et al. 2002, Auri\`ere et al. 2010, Makaganiuk et
al. 2011) as well as deep spectropolarimetric runs
dedicated to specific objects (e.g. Wade et al. 2006, Kochukhov et al. 2011), have ruled out the existence of such a population. 
These non-detections instead 
revealed a gap in longitudinal field between 100 G, the lower bound of Ap/Bp magnetism,
and the
detection limit of the surveys.
With MuSiCoS, \cite{SW02} reached a detection 
limit of the order of 50 G for the longitudinal fields, while the Narval survey (Auri\`ere et al. 2010) set a $3 \sigma$ upper limit of the longitudinal field to 10 G for most stars of the sample
down to $\sim$ 1 G in some bright low v sini targets like Sirius.
In the following, we shall speak of the magnetic desert in longitudinal field to describe this gap.

Its discovery among intermediate-mass stars has important consequences :
The first one is that Ap/Bp stars constitute a separated class of stars in what concerns their magnetic properties because 
no low field continuous extension of this large-scale stable magnetism could be found among intermediate-mass stars. 
The so-called 10\% problem of why
only a small fraction of stars appears to be magnetic 
is therefore not due to an observational bias but
corresponds instead to a true 
physical dichotomy between Ap magnetism and other intermediate-mass stars.
The lower bound of Ap/Bp star magnetic fields must thus be regarded as a characteristic property of Ap/Bp magnetism
whose
dependency on the star fundamental parameters should provide useful insight into the origin of this magnetism.
We shall come back to this point in section 3.


\begin{figure}[htb]
\resizebox{\hsize}{!}{\includegraphics{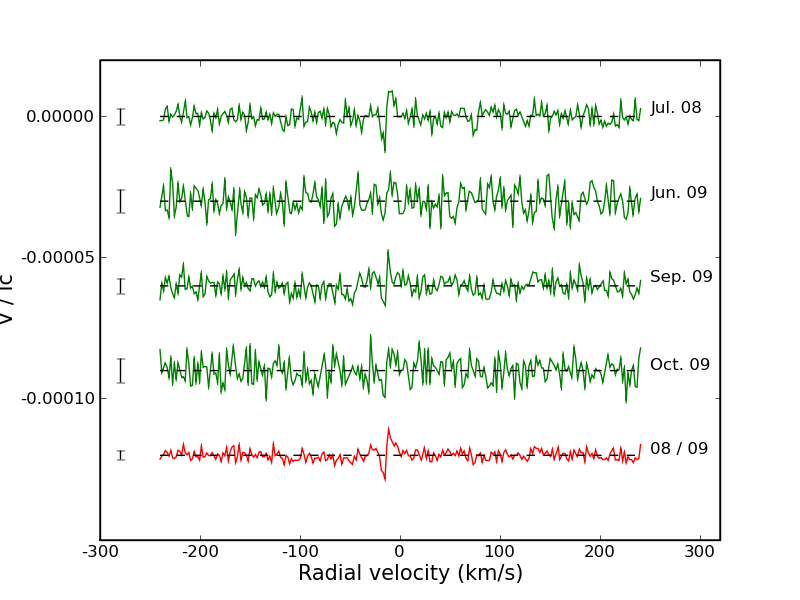}}
\caption{(Colour online) Averaged Stokes V LSD profiles of Vega for 4 different observing runs from July 2008 to October 2009.
The red profile is obtained by averaging all 799 spectra taken over this period.}
\label{fig0}
\end{figure}

\begin{figure}[htb]
\resizebox{\hsize}{!}{\includegraphics{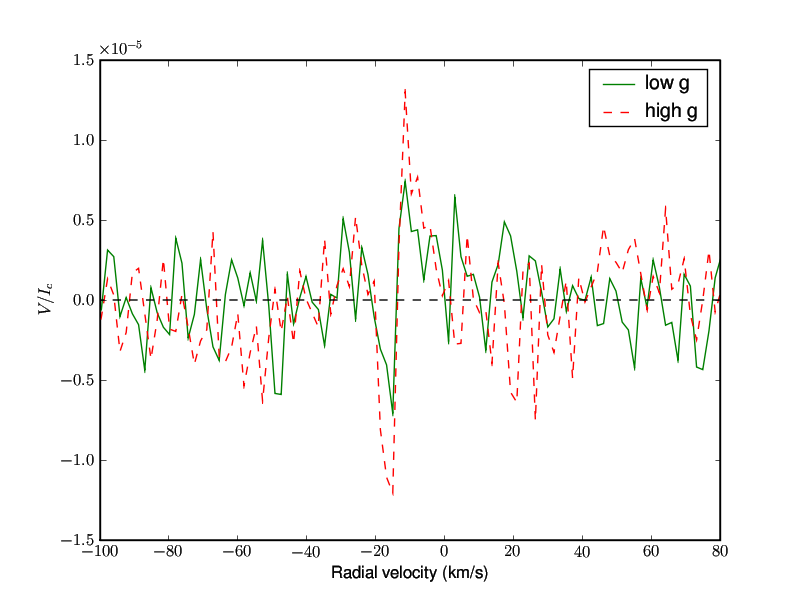}}
\caption{(Colour online) Averaged Stokes V LSD profiles of Vega for two different lines lists. The green and solid profile corresponds
to the line list with low Land\'e factor, the red and dashed profile to high Land\'e factor. As expected for a Zeeman signal, 
the amplitude ratio between the two Stokes V profiles is close to the 1.6 ratio between the mean Land\'e factor of the line lists.}
\label{fig1}
\end{figure}

\section{Ultra-weak magnetic field}

\begin{figure}[htb]
\resizebox{0.8\hsize}{!}{\includegraphics{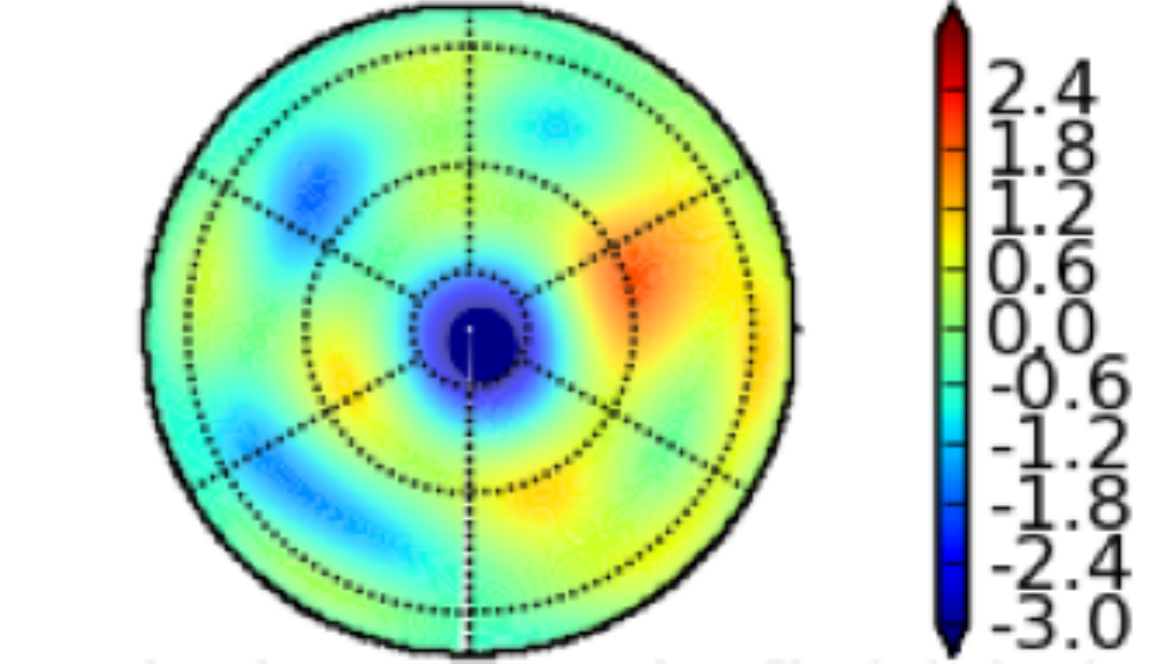}}	
\caption{(Colour online) Zeeman-Doppler Imaging of Vega for 2008 July showing the radial component of the magnetic field in polar projection. The color code is expressed in gauss. The spot observed near the pole can be related to the near-zero Doppler velocity feature of the Stokes V profile.}
\label{fig2}
\end{figure}

While revealing 
a magnetic desert among intermediate-mass stars,
the spectropolarimetric surveys mentioned in the previous section 
could only provide 
an upper bound of the surface magnetic field
of A-type non-Ap/Bp stars.
Fortunately, a 4-nights Narval run designed to search for pulsations on the A0 star Vega was used as an 
opportunity to further lower this bound and resulted in the detection of a very small polarimetric signal. The peak Stokes V amplitude
divided by the continuum intensity 
was $V/I_c=10^{-5}$, for a noise level of $\sigma = 2 \times 10^{-6}$. 
The corresponding 
longitudinal magnetic field is -0.6 $\pm$ 0.3 G (Ligni\`eres et al. 2009).
This weak signal called for additional measurements and tests to confirm the presence of 
a magnetic field.
As illustrated on Figure 1, additional observing campaigns conducted with Narval and ESPaDOnS over three years all confirmed
this first detection (Petit et al. 2010, Alina et al. 2012).
The possibility of a spurious detection 
due to the instrument or
the reduction process is systematically tested by computing the combination
of sub-exposures, the so-called null profiles, for which the signal is expected to cancel.
On the other hand, if the signal is of stellar origin, it should be sensitive to the Land\'e 
factor of the spectral lines as well as periodically modulated by the star rotation.
Figure 2 shows the Stokes V LSD profiles computed for two lines lists having different mean Land\'e factors. 
As expected for a Zeeman signature,
the amplitude ratio between the two Stokes V profiles is close to the 1.6 ratio of their mean Land\'e factor.
A periodic modulation at 0.678 $\pm$ 0.036 day has also been detected by applying a least squares sine fit period search to each radial 
velocity bin of
the Stokes V LSD profile (Alina et al. 2012).
This value can be compared with two different determinations of the rotation rate that both make use of the gravity darkening effect, that is the effect of the centrifugal force
on the local atmospheric parameters. The line shapes fitting by Takeda et al. (2008) provided 0.7-0.9 day for the rotation period  together with a small $7$ degrees inclination angle. By contrast the first model fitting of the interferometric observations gave much higher close-to-break-up periods (0.5-0.6 days)
with nevertheless similar inclination
angles. In a recent study however,  Monnier et al. (2012) showed that the error bars of these
interferometric determinations were underestimated and a more accurate value now
consistent with spectropolarimetric and spectroscopic inferences was obtained thanks to better phase coverages and angular resolutions.
Accordingly, the equatorial velocity is 194 $\pm 6$  km s$^{-1}$, a value only slightly above the average equatorial velocity of normal A0-A1 stars v$_{\rm eq}$ = 165 km s$^{-1}$ 
taken from
Royer et al. (2007).

Finding the rotation period is key to reconstruct the surface magnetic field through
Zeeman-Doppler Imaging.
The magnetic field vector appears to be dominated by its radial component whose surface distribution is structured at small lengthscale and
characterized by a slightly off-axis polar spot, as seen on Figure 3. 
This polar spot can already be inferred from the phase-averaged Stokes V concentration in the weakly Doppler-shifted part of the profile.

If we exclude the debated field detections obtained with the low-resolution FORS1/2 instrument at VLT
(e.g. Bagnulo et al. 2012, Kochukhov 2013),
Vega is the first detected magnetic A star which is not an Ap/Bp chemically peculiar star.
Its magnetic field distinguishes
clearly from Ap/Bp magnetic fields by the strength of the longitudinal component (two orders
of magnetic lower than the minimum longitudinal field of Ap/Bp stars taken at its maximum phase) and
by the lengthscale of its surface distribution (significantly smaller than for a dipole).
It shows that  a new - and potentially widespread - class of magnetic stars exists
at the low end of the A-type magnetic desert.


The next obvious target is Sirius A, a low v sini bright A1 star with Am-type chemical abundance anomalies.
The LSD Stokes V profile obtained by Petit et al. (2011) from the average of 442 high S/N ratio spectra clearly shows a polarimetric signal.
The $V/I_c$ maximum amplitude is similar to Vega but, contrary to Vega, the V profile is asymmetric about the line center. Asymmetric profiles are 
not standard although they are observed in some cool stars (Petit et al. 2005, Auri\`ere et al. 2009) and locally on the Sun surface where they are 
interpreted as 
due to the presence of both magnetic field and velocity field radial gradients 
in the emitting regions (Lopez-Ariste 2002). A model of
stellar atmosphere that produces an asymmetric profile similar to Sirius on a global scale is still lacking.
Moreover, the quality of the data has not allowed yet to show a significant sensitivity to the Land\'e
factor or a temporal modulation of the polarimetric signal.
Thus, by contrast with Vega, more theoretical and observational works are required to confirm the presence of 
a magnetic field on Sirius.
A spurious detection seems however unlikely as the V profile found with ESPaDOnS and Narval has been recently
recovered using HARPSpol (see Kochukhov 2013).
The rotation of Sirius is not known but as the equatorial velocities of Am stars are always smaller than
120 km s$^{-1}$ (Abt 2009) we know that Sirius rotates more slowly than a typical A0-A1 star.


An ongoing  Narval survey investigates the occurrence of Vega-like magnetism among tepid stars.
The weakness
of the expected signal, the necessity to avoid confusion with hypothetical weak solar-type dynamo field in colder stars, the drastic
decrease in the number of lines towards hotter stars and the willingness to avoid a bias towards the slowly rotating Am stars
contributed to construct a sample of 10 bright stars restricted to the A0-A2 range (including Vega and Sirius).
This survey also intends to investigate how Vega-like magnetism depends on rotation and time.
A large occurrence of magnetic field detections would have a direct impact on stellar evolution model
by providing the first direct constraint on the value of the magnetic field of a typical intermediate-mass star.
Meanwhile, the hypothesis of a widespread magnetism is suggested by the analysis of the Kepler photometry of
thousands of A stars.
Indeed, according to Balona (2011), a low frequency modulation of the
light-curve compatible with a rotational modulation has been found in 70 \% of the A-type Kepler stars, as expected
in the 
presence of starspots or other 
magnetic corotating features.


\begin{figure}[htb]
\resizebox{0.8\hsize}{!}{\includegraphics{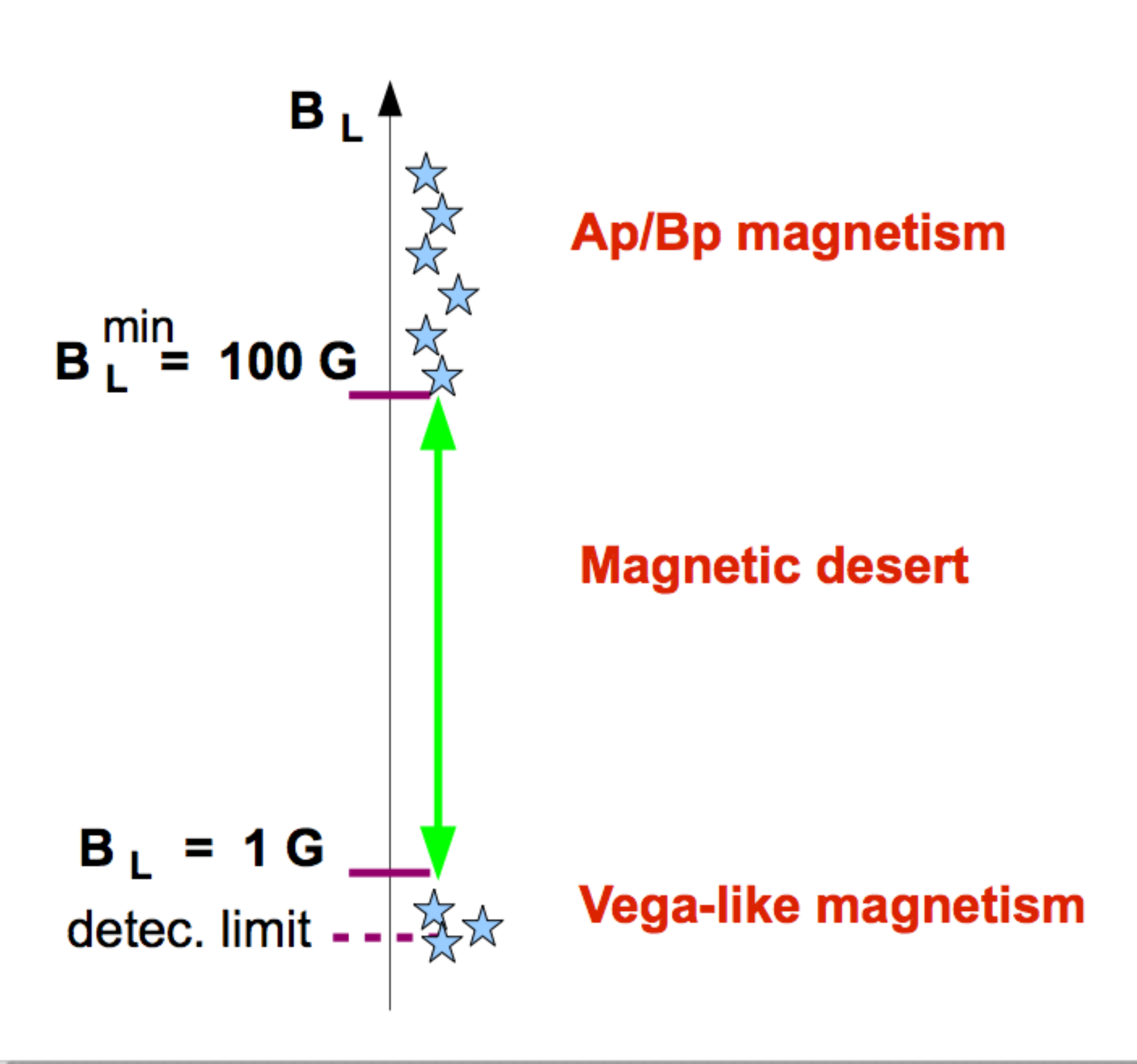}}
\caption{(Colour online) This sketch summarizes the new context	of intermediate-mass magnetism set by the recent discoveries 
of the lower bound of Ap magnetism, the magnetic desert in longitudinal field $B_L$ and the Vega-like magnetism. }
\label{fig3}
\end{figure}

\begin{figure}[htb]
\resizebox{\hsize}{!}{\includegraphics{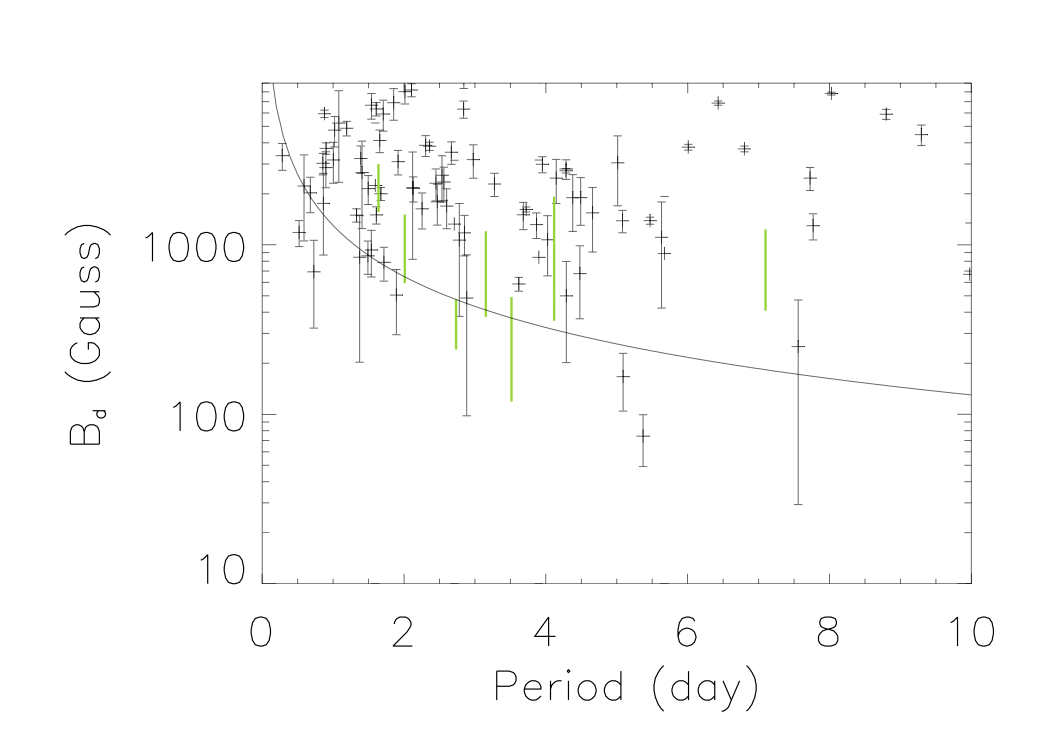}}
\caption{(Colour online) Ap/Bp dipolar field strengths
as a function of the rotation rate.
The black vertical line segments are 
dipole strength (with error bars) derived from the longitudinal field
measurements of the Bychkov's catalogue and the inequality : $B_d > 3.3 B_L^{\mbox{max}}$.
The green segments correspond to dipole strength (with error bars) 
obtained by fitting an oblique rotator model (taken from Auri\`ere et al. 2007). 
The black continuous line is a $B_d \propto \Omega$ curve.}

\label{fig3}
\end{figure}

\section{On the origin of the magnetic dichotomy}

The sketch of Fig. 4 illustrates the dichotomy between the Ap/Bp-like and Vega-like magnetic fields
of intermediate-mass stars. One can think of two different ways to explain this dichotomy :
the first one is to assume that the two types of magnetic fields have different properties because they have been generated by two different processes.
Another possibility is that the observed fields have a common origin but during the evolution their magnetic field distribution split into two distinct families
of low and high longitudinal fields.

Braithwaite \& Cantiello (2013) proposed that  Vega-like stars are "failed fossil" magnetic stars meaning that
their field, produced during star formation, is still decaying thus not truly fossil. 
Accordingly, the helicity of the initial field configuration must be low enough to evolve towards
a sub-gauss 
field amplitude at the age of Vega. By contrast, the initial helicity of Ap/Bp-like magnetic stars has to be very high 
to produce in a relatively short time the observed fossil-like Ap/Bp magnetism.
Braithwaite \& Cantiello (2013) thus argue that two distinct generation mechanisms must be invoked to explain 
these very different initial helicities. To account for Ap/Bp stars, they follow Ferrario et al. (2009) and Tutukov\& Fedorova (2010)
who assume that these stars 
result
from the merging of close binaries, 
their strong fields being produced by a powerful dynamo during the merging phase.

On the other hand, \cite{AG07} proposed a bifurcation scenario by considering the evolution of a differentially rotating star
as a function of its rotation rate and its poloidal field strength. 
Weak initial poloidal fields cannot prevent their winding-up by differential rotation
into strong toroidal fields. The increasingly toroidal configuration is expected to
become unstable to an instability, like the Tayler instability, transforming
the initially large scale field configuration into a new configuration with opposed signed polarities at the length
scale of the instability. On the contrary, if the initial magnetic field is strong enough,
Maxwell stresses impose uniform rotation and eventually lead to a stable
configuration. Thus, starting from a continuous distribution of poloidal field strength 
ranging from low
to high dipole strengths,
this mechanism predicts a sharp decrease of the longitudinal
field between the stable large-scale high B$_L$ fields and the unstable configurations where B$_L$, the surface averaged line-of-sight component, is strongly reduced due to the cancellation of the opposed polarities.
In support of this scenario, 
the observed $\sim 300$ G lower bound of Ap/Bp magnetic fields (see section 2) happens to be close to an order of magnitude estimate of the critical dipolar field
separating the stable and unstable configuration, namely $B_c = (4 \pi \rho)^{1/2} r \Omega$, computed for
the parameters of a typical Ap star ($\log g=4 , T_{\rm eff} = 10^4 K, P_{\rm rot}=5$ days).

Another important property of B$_c$ is to be proportional to $\Omega$. Thus if this scenario can explain the dichotomy, the lower bound of Ap/Bp magnetic 
fields should increase with $\Omega$. This can be tested since the spectropolarimetric monitoring of an Ap/Bp star gives access to both
the rotation period and the dipole strength. The difficulty is that finding the minimum dipolar fields of Ap/Bp stars for various rotation rate intervals 
requires a significant sample of Ap/Bp stars for each rotation range.
As a first attempt, Auri\`ere \& Ligni\`eres (2013) compiled data from the Bychkov
catalogue (Bychkov et al. 2005) although in many cases the data are not good
enough for an oblique rotator modelling. But only with the maximum value of the longitudinal field
B$_L^{\mbox{max}}$ available, a lower limit of the dipole strength can be determined from 
the relation $B_d > 3.3 B_L^{\mbox{max}}$ derived from the oblique rotator model.
These lower limits are displayed on Fig. 5 with errors bars together with $ B_d(\Omega)$ values taken from  Auri\`ere et al. (2007).
The results seem compatible
with a B$_{\rm min} \propto \Omega$ law. Nevertheless this data sample is too 
limited 
to obtain an accurate determinations of B$_{\rm min}$, specially for rapid Ap/Bp rotators. This calls for
a comprehensive investigation of the  B$_{\rm min}$ dependency on the rotation rates. More generally, we should expect that 
this lower bound, and its dependency on the stellar parameters, contains important information about the origin of Ap/Bp magnetism.

The physical grounds of the Auri\`ere et al. scenario can also be investigated through 2D and 3D numerical simulations. The winding-up
of a poloidal field by differential rotation is accessible to 2D axisymmetric simulations while the non axisymmetric instabilities 
of the wounded-up configuration require full 3D simulations (e.g. Arlt \& R\"udiger 2011). Among the open questions is the mechanism that
drives 
differential rotation and whether this mechanism is efficient enough to overcome Maxwell stresses.
The seismic analysis of sub-giant stars has shown recently that differential rotation is indeed present
in the contracting/expanding radiative interiors of these stars (Deheuvels et al. 2012). This points towards pre-main-sequence contraction as a 
possible source for the differential rotation forcing required
in the Auri\`ere et al. scenario.

\section{Conclusion}
While some years ago the fraction of intermediate-mass stars hosting large-scale stable magnetic fields like Ap/Bp stars was unknown,
this class of magnetic stars is now known to be clearly separated from the other stars by a two order of magnitude magnetic desert in longitudinal field 
down to a new type of sub-gauss magnetism.
This paradigm shift opens new perspectives for both observational and theoretical studies. Investigating how the lower bound of Ap/Bp magnetic fields 
depends on the stellar parameters
could help to discriminate between the different hypothesis 
regarding the origin of this magnetism. For example we have seen that the Auri\`ere et al. scenario predicts that this
lower bound is proportional to the rotation rate.
Moreover, the nature and the incidence of Vega-like magnetism among non Ap/Bp stars need to be elucidated 
although detecting many sub-gauss fields in A stars will be very demanding in observing time.
Another important question is whether the dichotomy of intermediate-mass star magnetism is also
present among 
pre-main-sequence intermediate-mass stars and massive stars. This is already suggested by recent spectropolarimetric surveys
that detected fossil-like magnetic fields with an incidence similar to Ap/Bp stars (Alecian et al. 2013, Wade et al. 2013). 

Various scenarios have been proposed to account for the magnetic dichotomy. Here we described the bifurcation
between stable and unstable magnetic configurations and briefly mentioned 
the merging scenario for Ap/Bp stars and the failed fossil scenario for Vega-like magnetism (see the contributions by 
N. Langer and J. Braithwaite in these proceedings for more details). Besides 
the confrontation with observations, numerical simulations will also help to test
these ideas. MHD computations in polytropic stars already enabled to find stable
magnetic
configurations that might correspond to Ap/Bp stars (Braithwaite 2009). 
But, except for very strong fields,
differential rotation should also play an important role in shaping magnetic configurations 
in radiative interiors and this effect needs to be fully explored numerically.


\begin{acknowledgements}
The authors thank the ANR project IMAGINE and the PNPS (Programme National de Physique Stellaire)
of INSU
for their financial support.
\end{acknowledgements}


\begin{thebibliography}{}

\bibitem[Abt(2009)]{A09} Abt, H.~A.\ 2009, \textit{AJ}, 138, 28

\bibitem[Alecian(2013)]{A13} Alecian, E.\ 2013, 
arXiv:1310.1725 

\bibitem[Alina et al.(2012)]{AP12} Alina, D., Petit, P., 
Ligni{\`e}res, F., et al.\ 2012, \textit{AIP Conference 
Series}, 1429, 82

\bibitem[Arlt  
(2011)]{AR11} Arlt, R., R\"udiger, G.\ 2011, \textit{MNRAS}, 412, 107

\bibitem[Auri{\`e}re \& Ligni{\`e}res
(2013)]{AG10} Auri{\`e}re, M., \& Ligni{\`e}res, F.\ 2013, \textit{private communication}

\bibitem[Auri{\`e}re et 
al.(2010)]{AG10} Auri{\`e}re, M., Wade, G.~A., Ligni{\`e}res, F., et al.\ 2010, \textit{A\&A}, 523, A40

\bibitem[Auri\`ere \etal\ (2009)]{AK09}
Auri{\`e}re, M., Wade, G.~A., Konstantinova-Antova, R., et al.\ 2009, \textit{A\&A}, 504, 231

\bibitem[Auri{\`e}re et 
al.(2008)]{AK08} Auri{\`e}re, M., Konstantinova-Antova, R., Petit, P., et al.\ 2008, \textit{A\&A}, 491, 499 

\bibitem[Auri{\`e}re et 
al. (2007)]{AG07} Auri{\`e}re, M., Wade, G.~A., Silvester, J., et al.\ 2007, \textit{A\&A}, 475, 1053 

\bibitem[Babcock (1947)]{B47} Babcock, H.~W.\ 1947, \textit{ApJ}, 
105, 105

\bibitem[Bagnulo et 
al.(2012)]{BL12} Bagnulo, S., Landstreet, J.~D., Fossati, L., \& Kochukhov, O.\ 2012, \textit{A\&A}, 538, A129

\bibitem[Balona (2011)]{B11} Balona, L.~A.\ 2011, \textit{MNRAS},
415, 1691

\bibitem[Braithwaite 
\& Cantiello(2013)]{BC13} Braithwaite, J., \& Cantiello, M.\ 2013, \textit{MNRAS}, 428, 2789

\bibitem[Braithwaite(2009)]{B09} Braithwaite, J.\ 2009, 
\textit{MNRAS}, 397, 763

\bibitem[Bychkov et 
al.(2005)]{BB05} Bychkov, V.~D., Bychkova, L.~V., \& Madej, J.\ 2005, \textit{A\&A}, 430, 1143

\bibitem[Donati 
\& Landstreet(2009)]{2009ARA&A..47..333D} Donati, J.-F., \& Landstreet, J.~D.\ 2009, \textit{ARA\&A}, 47, 333

\bibitem[Deheuvels et al.(2012)]{DG12} Deheuvels, S., 
Garc{\'{\i}}a, R.~A., Chaplin, W.~J., et al.\ 2012, textit{ApJ}, 756, 19

\bibitem[Ferrario et al.(2009)]{FP09} Ferrario, L., Pringle, 
J.~E., Tout, C.~A., \& Wickramasinghe, D.~T.\ 2009, \textit{MNRAS}, 400, L71

\bibitem[Kochukhov et 
al.(2011)]{KM11} Kochukhov, O., Makaganiuk, V., Piskunov, N., et al.\ 2011, \textit{A\&A}, 534, L13

\bibitem[Kochukhov(2013)]{K13} Kochukhov, O.\ 2013, 
arXiv:1309.6313

\bibitem[Landstreet(1982)]{L82} Landstreet, J.~D.\ 1982, 
\textit{ApJ}, 258, 639

\bibitem[Ligni{\`e}res et 
al.(2009)]{LP09} Ligni{\`e}res, F., Petit, P., B{\"o}hm, T., \& Auri{\`e}re, M.\ 2009, \textit{A\&A}, 500, L41

\bibitem[L{\'o}pez Ariste(2002)]{L02} L{\'o}pez Ariste, A.\ 
2002, \textit{ApJ}, 564, 379

\bibitem[Makaganiuk et 
al.(2011)]{MK11} Makaganiuk, V., Kochukhov, O., Piskunov, N., et al.\ 2011, \textit{A\&A}, 525, A97

\bibitem[Michaud et al.(1976)]{MC76} Michaud, G., Charland, 
Y., Vauclair, S., \& Vauclair, G.\ 1976, \textit{ApJ}, 210, 447

\bibitem[Michaud(1970)]{M70} Michaud, G.\ 1970, \textit{ApJ}, 160,
641

\bibitem[Monnier et al.(2012)]{MC012} Monnier, J.~D., Che, 
X., Zhao, M., et al.\ 2012, \textit{ApJL}, 761, L3

\bibitem[Petit et 
al.(2011)]{PL11} Petit, P., Ligni{\`e}res, F., Auri{\`e}re, M., et al.\ 2011, \textit{A\&A}, 532, L13

\bibitem[Petit et 
al.(2010)]{PL10} Petit, P., Ligni{\`e}res, F., Wade, G.~A., et al.\ 2010, \textit{A\&A}, 523, A41

\bibitem[Petit et al.(2005)]{PD05} Petit, P., Donati, J.-F., 
Auri{\`e}re, M., et al.\ 2005, \textit{MNRAS}, 361, 837 

\bibitem[Royer et 
al.(2007)]{RZ07} Royer, F., Zorec, J., \& G{\'o}mez, A.~E.\ 2007, \textit{A\&A}, 463, 671

\bibitem[Shorlin et 
al.(2002)]{SW02} Shorlin, S.~L.~S., Wade, G.~A., Donati, J.-F., et al.\ 2002, \textit{A\&A}, 392, 637

\bibitem[Takeda et al.(2008)]{TK08} Takeda, Y., Kawanomoto, 
S., \& Ohishi, N.\ 2008, \textit{ApJ}, 678, 446

\bibitem[Tutukov 
\& Fedorova(2010)]{TF10} Tutukov, A.~V., \& Fedorova, A.~V.\ 2010, \textit{Astronomy Reports}, 54, 156

\bibitem[Wade et al.(2013)]{WG13} Wade, G.~A., Grunhut, J., 
Alecian, E., et al.\ 2013, arXiv:1310.3965

\bibitem[Wade et 
al.(2006)]{WA06} Wade, G.~A., Auri{\`e}re, M., Bagnulo, S., et al.\ 2006, \textit{A\&A}, 451, 293

\end{thebibliography}
\end{document}